\newcommand{\be}{\begin{equation}}
\newcommand{\ee}{\end{equation}}
\newcommand{\bea}{\begin{eqnarray}}
\newcommand{\eea}{\end{eqnarray}}

\newcommand{\aeq}{&=&}

\newcommand{\itDelta}{{\it \Delta}}

\newcommand{\itLambda}{{\Lambda}}

\newcommand{\bra}{\langle}
\newcommand{\ket}{\rangle}
\newcommand{\dbra}{\bra \! \bra}
\newcommand{\dket}{\ket \! \ket}

\newcommand{\me}{\mbox{e}}

\newcommand{\bq}{{\bar q}}

\newcommand{\rS}{{\rm S}}

\documentclass{elsart}
\usepackage{epsfig}

\usepackage{amssymb}

\begin{document}

\begin{frontmatter}



\title{Harmonious Representation of PDF's 
reflecting Large Deviations
}




\author[label2]{Toshihico~Arimitsu}\ead{arimitsu@cm.ph.tsukuba.ac.jp} and
\author[label3]{Naoko~Arimitsu}

\address[label2]{Institute of Physics, University of Tsukuba, Ibaraki 305-8571,
Japan}
\address[label3]{Graduate School of EIS, Yokohama Nat'l.~University, 
Yokohama 240-8501, Japan}

\begin{abstract}
The framework of multifractal analysis (MFA) is distilled
to the most sophisticated one. Within this transparent framework, 
it is shown that 
the harmonious representation of MFA utilizing two distinct 
Tsallis distribution functions,
one for the tail part of probability density function (PDF) 
and the other for its center part, 
explains the recently observed PDF's of turbulence
in the highest accuracy superior to the analyses based on other models such as 
the log-normal model and the $p$ model.
\end{abstract}

\end{frontmatter}

\section{Introduction}
\label{intro}

The {\it multifractal analysis} (MFA) is a unified self-consistent approach 
dealing with those systems exhibiting large deviations.
It has been proposed by the present authors \cite{AA,AA1,AA4,AA5,AA7,AA8,AA11,AA10,AA12} 
in order to analyze various probability density functions (PDF's) 
observed in the system of fully developed turbulence.
The PDF's, given for variables normalized by their own standard deviations, 
represent a characteristic large wings (or fat tails) 
due to the intermittency specific to turbulence.
Since the pioneering work by Kolmogorov \cite{K41}, there appeared so many 
attractive attempts to analyze the large wings, 
e.g., the log-normal model \cite{Oboukhov62,K62,Yaglom},
the $p$ model \cite{Meneveau87a,Meneveau87b} etc., 
but no one was satisfactory.
The MFA is constructed with the help of 
the Tsallis-type distribution function \cite{Tsallis88}
that provides an extremum of the {\it extensive} R\'{e}nyi entropy \cite{Renyi} 
or the {\it non-extensive} Tsallis entropy \cite{Tsallis88,Havrda-Charvat}
under appropriate constraints \cite{AA4}.
After a rather preliminary investigation of the $p$ model \cite{AA},
we developed further to derive the analytical expression for 
the scaling exponents of velocity structure function \cite{AA1,AA4}, and to
determine the PDF's of the velocity fluctuations 
\cite{AA4,AA5,AA7}, of the velocity derivative \cite{AA8} and of the fluid particle
accelerations \cite{AA11}.

In this paper, we will show that a harmonious representation of PDF's 
by means of two distinct Tsallis-type distributions provides us with a description for 
experimentally or simulationally observed PDF's in the highest accuracy
compared with other multifractal model such as 
the log-normal model \cite{Oboukhov62,K62,Yaglom}
and the $p$ model \cite{Meneveau87a,Meneveau87b}.
In order to perform a transparent comparison, the framework of MFA is 
rephrased in its most sophisticated fashion in which 
the {\em tail part} of PDF, giving the probabilities of events 
larger than its standard deviation, is written down once the multifractal spectrum 
for the spatial distribution of singularities is specified
(see (\ref{PDF large phi}) below),
whereas the {\em center part}, giving the probabilities of events
smaller than its standard deviation, is assumed to be analyzed by 
the Tsallis-type PDF for the variable itself \cite{AA10,AA12}
(see (\ref{PDF small phi}) below).
The various PDF's extracted out from the DNS conducted by Gotoh et al.\ 
\cite{Gotoh02,Gotoh pressure}, and the PDF of fluid particle accelerations 
observed by Bodenschatz et al.\ in their Lagrangian measurement of particle accelerations 
\cite{EB01a,EB01b,EB02comment} will be analyzed 
with the help of the multifractal spectrums for the log-normal, the $p$ model
and the harmonious representation.

MFA of turbulence rests
on the scale invariance of the Navier-Stokes equation for high Reynolds number, 
and on the assumption that the singularities due to the invariance
distribute themselves, multifractally, in physical space.
Let us consider the fluctuation 
$
\delta x_n = \vert x(\bullet + \ell_n) - x(\bullet) \vert
$
of a physical quantity $x$ at the $n$th multifractal step 
satisfying the scaling law
$
\vert x_n \vert \equiv \left\vert \delta x_n/\delta x_0 \right\vert 
= \delta_n^{\ \alpha \phi/3}
$
with
$
\delta_n = \ell_n /\ell_0 = \delta^{-n}
\label{r-n}
$
$
(n=0,1,2,\cdots)
$.
We call $n$ the multifractal depth which can be real number in the analysis of 
experimental data.
In the following in this paper, we will put $\delta =2$ that is consistent with
the energy cascade model.
At each step of the cascade, say at the $n$th step, eddies break up into 
two pieces producing the energy cascade with the energy-transfer rate
$\epsilon_n$ that represents the rate of transfer of energy per unit mass 
from eddies with diameter $\ell_n$ to those with $\ell_{n+1}$.
Then, we see that the derivative 
$
\vert x^\prime \vert = \lim_{n \rightarrow \infty} x^\prime_n
$
with the $n$th difference
$
x'_n = \delta x_n/\ell_n
$
for the characteristic length $\ell_n$ diverges for $\alpha < 3/\phi$ \cite{Benzi84}.
The velocity fluctuation $\delta u_n$ is given for $\phi = 1$, 
and the pressure fluctuation $\delta p_n$ is for $\phi = 2$.
The velocity derivative and the fluid particle acceleration are defined, respectively, by
$
\vert u^\prime \vert = \lim_{n \rightarrow \infty} u^\prime_n
$
and 
$
\vert \vec{\mathrm{a}} \vert = \lim_{n \rightarrow \infty} \mathrm{a}_n
$
with the $n$th velocity difference 
$
u'_n = \delta u_n/\ell_n
$
and the $n$th pressure difference (the $n$th acceleration)
$
{\mathrm{a}}_n = \delta p_n/\ell_n\
$.
Note that the fluid particle acceleration $\vec {\mathrm{a}}$ is
given by 
$
{\vec {\mathrm a}} = \partial {\vec u}/\partial t
+ ( {\vec u}\cdot {\vec \nabla} ) {\vec u}
$.
For the energy transfer rate, $\phi = 3$ since it has the scaling relation
$
\epsilon_n/\epsilon_0 = \delta_n^{\ \alpha -1}
$.

The real quantity $\alpha$ is introduced 
in the scale transformation \cite{Moiseev76,Frisch-Parisi83}
$
{\vec x} \rightarrow {\vec x}'=\lambda {\vec x},\ 
{\vec u} \rightarrow {\vec u}'=\lambda^{\alpha/3} {\vec u},\ 
t \rightarrow t'=\lambda^{1- \alpha/3} t,\ 
p \rightarrow p'=\lambda^{2\alpha/3} p
\label{scale trans}
$
that leaves the Navier-Stokes equation 
$
\partial {\vec u}/\partial t
+ ( {\vec u}\cdot {\vec \nabla} ) {\vec u} 
= - {\vec \nabla} p
+ \nu \nabla^2 {\vec u}
\label{N-S eq}
$
of incompressible fluid invariant when the Reynolds number 
$
{\rm Re}=\delta u_{\rm in} \ell_{\rm in}/\nu
$ is large.
Here, $\nu$ is the kinematic viscosity, and $p=\check{p}/\rho$ with
the thermodynamical pressure $\check{p}$ and the mass density $\rho$. 
$\delta u_{\rm in}$ and $\ell_{\rm in}$ represent, respectively, 
the rotating velocity and the diameter of the largest eddies in turbulence.
The largest size of eddies is, for example, about the order of mesh size of a grid, 
inserted in a laminar flow, that produces turbulence downstream.

\section{General framework}

MFA starts with an assignment of the probability, to find a singularity 
specified by the strength $\alpha$ within the range
$
\alpha \sim \alpha + d \alpha
$,
in the form
$
P^{(n)}(\alpha) d\alpha = (\vert f^{\prime \prime}(\alpha_0) \vert
\vert \ln \delta_n \vert /2\pi)^{1/2}\ \delta_n^{1-f(\alpha)}\ d\alpha
\label{P alpha f}
$ 
\cite{Meneveau87b,AA4}.
Here, $f(\alpha)$ represents an appropriate multifractal spectrum 
defined in the range $\alpha_{\rm min} \leq \alpha \leq \alpha_{\rm max}$.
Note that $f(\alpha)$ does not dependent on $n$ because of the scale invariance.
The multifractal spectrum is related to the mass exponent $\tau(\bar{q})$, defined by
$
\bra (\epsilon_n/\epsilon )^{\bar{q}} \ket
= a_{3\bq}\ \delta_n^{-\tau(\bar{q}) +1 -\bar{q}}
$
with
$
a_{3\bq} = (\vert f^{\prime \prime}(\alpha_0) \vert/
\vert f^{\prime \prime}(\alpha_{\bq}) \vert)^{1/2}
$,
through the Legendre transformation \cite{Meneveau87b}:
$
f(\alpha) = \alpha \bq + \tau(\bq)
\label{f-tau1}
$
with
$
\alpha = \alpha_{\bq} = - d \tau(\bq)/d \bq
\label{alpha-tau1}
$
and
$
\bq = d f(\alpha)/d \alpha
\label{def of bq1}
$.
The average $\bra \cdots \ket$ is taken with $P^{(n)}(\alpha)$, and
$\alpha_0 = \alpha_{\bq=0} = \bra \alpha \ket$.

The PDF
$
\Pi_\phi^{(n)}(x_n) 
$, 
normalized as
$
\int dx_n \Pi_\phi^{(n)}(x_n) = 1
$,
is assumed to consists of two parts, i.e.,
$
\Pi_\phi^{(n)}(x_n) = \Pi^{(n)}_{\phi,\rS}(x_n) + \Delta \Pi_\phi^{(n)}(x_n) 
\label{def of Pi phi}
$
where the first term is related to $P^{(n)}(\alpha)$ through
$
\Pi^{(n)}_{\phi,\rS}(\vert x_n \vert) d \vert x_n \vert 
= [(1-2\gamma_{\phi,0}^{(n)})/2] P^{(n)}(\alpha) d \alpha
\label{singular portion}
$
with the transformation of the variables 
$
\vert x_n \vert = \delta_n^{\ \alpha \phi/3}
$, 
and the second term is responsible to the contributions coming from 
the dissipative term in the Navier-Stokes equation violating the invariance under 
the scale transformation given above.
Here, we introduced $\gamma^{(n)}_{\phi,m}$ by
$
2\gamma^{(n)}_{\phi,m} = \int dx_n 
\vert x_n \vert^m \Delta \Pi_\phi^{(n)}(x_n)
$.
Note that we are dealing with the symmetrized part of PDF's by assuming that 
the large deviation stemmed from the singular first term contributes to symmetric part
of PDF's.
We have the $m$th order structure function for $x_n$ in the form
$
\dbra \vert x_n \vert^m \dket_\phi \equiv \int dx_n  
\vert x_n \vert^m \Pi_\phi^{(n)}(x_n)
= 2 \gamma^{(n)}_{\phi,m}
+ (1-2\gamma^{(n)}_{\phi,0} ) a_{\phi m} \delta_n^{\ \zeta_{\phi m}}
\label{structure func m}
$
with the corresponding scaling exponents
$
\zeta_{3\bq} = 1-\tau(\bq)
$.
For $\phi = 1$, $\zeta_{\phi m}$ reduces to the scaling exponents $\zeta_m$ 
of the $m$th order velocity structure function.

Let us introduce the PDF 
$
\hat{\Pi}_\phi^{(n)}(\xi_n)
$,
to be compared with observed data, defined by
$
\hat{\Pi}_\phi^{(n)}(\xi_n) d\xi_n = \Pi_\phi^{(n)}(x_n) d x_n
$
or by
$
\hat{\Pi}_\phi^{(n)}(\xi_n) d\xi_n = \Pi_\phi^{(n)}(x_n^\prime) d x_n^\prime
$
for the variable
$
\xi_n = x_n/\dbra x_n^2 \dket^{1/2} = x_n^\prime/\dbra (x_n^\prime)^2 \dket^{1/2} 
$
both of the fluctuation $x_n$ and of the derivative $x_n^\prime$ normalized by
their own standard deviations.
For values of the variable $\vert \xi_n \vert$ larger than the order of its standard deviation, 
$\xi_n^* \leq \vert \xi_n \vert \leq \xi_n^{\rm max}$ 
(equivalently, $\alpha_{\rm min} \leq \vert \alpha \vert \leq \alpha^*$), 
the PDF is given by \cite{AA10,AA12}
\be
\hat{\Pi}_\phi^{(n)}(\vert \xi_n \vert) d \vert \xi_n \vert
= \Pi^{(n)}_{\phi,\rm S} (\vert x_n \vert) d \vert x_n \vert
= [(1-2\gamma_{\phi,0}^{(n)})/2] P^{(n)}(\alpha) d\alpha
\label{PDF large phi}
\ee
where
$
\xi_n^{\rm max} = \bar{\xi}_n \delta_n^{\ \phi \alpha_{\rm min}/3 -\zeta_{2\phi} /2}
$
with 
$
\bar{\xi}_n = [2 \gamma_{\phi,2}^{(n)} \delta_n^{-\zeta_{2\phi}} 
+ (1-2\gamma_{\phi,0}^{(n)} ) a_{2\phi} ]^{-1/2}
$.
Note that $\xi_n^* \sim 1$ as can be seen below when we analyze experiments.
This {\em tail part} represents the large deviations, and manifests itself 
the multifractal distribution of the singularities due to the scale invariance 
of the Navier-Stokes equation when its dissipative term can be neglected.
The entropy index $q$ should be unique once a turbulent system with 
a certain Reynolds number is specified.
For smaller values of the variable, 
$
\vert \xi_n \vert \leq \xi_n^*
$ (equivalently, $\alpha^* \leq \vert \alpha \vert$),
we assume that the PDF has the Tsallis-type structure 
with a new entropy index $q'$ \cite{AA10,AA12}
\bea
\lefteqn{\hat{\Pi}_\phi^{(n)}(\xi_n) d \xi_n =
\left[ \Pi^{(n)}_{\phi,\rS}(x_n)
+\Delta \Pi_\phi^{(n)}(x_n) \right] d x_n}
\nonumber\\
\aeq
\bar{\Pi}_\phi^{(n)} \left\{1-(1-q') \left(\phi+3f'(\alpha^*)\right)
\left[ \left(\xi_n/\xi_n^* \right)^2 -1 \right]/2\phi \right\}^{1/(1-q')} d \xi_n
\label{PDF small phi}
\eea
where 
$
\bar{\Pi}_\phi^{(n)} = 3(1-2\gamma_{\phi,0}^{(n)}) 
\vert f^{\prime \prime}(\alpha_0) \vert^{1/2}/2\phi \bar{\xi}_n (2\pi
\vert \ln \delta_n\vert)^{1/2}
$.
This {\em center part} is responsible to smaller fluctuations of the variable, 
compared with its standard deviation, stemmed from the dissipative term violating 
the scale invariance.
The entropy index $q'$ can be dependent on the distance of two measuring points.

The two parts of the PDF, (\ref{PDF large phi}) and (\ref{PDF small phi}), are
connected at 
$
\xi_n^* = \bar{\xi}_n \delta_n^{\ \phi \alpha^* /3 -\zeta_{2\phi} /2}
$
under the conditions that they have the common value, $\bar{\Pi}_\phi^{(n)}$,
and the common log-slope, $-(\phi + 3f^\prime(\alpha^*))/\phi \xi_n^*$ there.
The value $\alpha^*$ is the smaller solution of
$
\zeta_{2\phi}/2 - \phi \alpha/3 +1 -f(\alpha) = 0
$.
The point $\xi_n^*$ has the characteristics that the dependence of
$
\hat{\Pi}_\phi^{(n)}(\xi_n^*)
$
on $n$ is minimum for large $n$ (see Fig.~4 in \cite{AA5}).
With the help of (\ref{PDF large phi}) and (\ref{PDF small phi}),
we obtain $\Delta \Pi_\phi^{(n)}(x_n)$, and have the analytical formula
$
2\gamma_{\phi,m}^{(n)} = (K_{\phi,m}^{(n)} - L_{\phi,m}^{(n)})/
(1+K_{\phi,0}^{(n)} - L_{\phi,0}^{(n)})
$
where
\bea
K_{\phi,m}^{(n)} \aeq \left(3\delta_n^{(m+1)\phi \alpha^*/3 -\zeta_{2\phi}/2}/\phi\right)
\sqrt{\vert f^{\prime \prime}(\alpha_0) \vert/2\pi \vert \ln \delta_n \vert} 
\nonumber\\
&& \times \int_0^1 dz\ \vert z \vert^m \left[1-(1-q^\prime)\left(\phi + 3f^\prime(\alpha^*)\right)
(z^2-1)/2\phi \right]^{1/(1-q^\prime)} 
\\
L_{\phi,m}^{(n)} \aeq \delta_n \sqrt{\vert f^{\prime \prime}(\alpha_0) \vert
\vert \ln \delta_n \vert/2\pi} 
\int_{\alpha^*}^{\alpha_{\rm max}} d\alpha \ \delta_n^{\ m\alpha \phi/3 - f(\alpha)}.
\eea
For later convenience, we introduce here the quantity
$
\xi_{n,0} = \bar{\xi}_n \delta_n^{\ \phi \alpha_0/3 -\zeta_{2\phi} /2}
$.

The PDF's $\hat{\Pi}^{(n)}(\xi_n)$ 
of the velocity fluctuations and of the velocity derivatives are give by
the common formula
$
\hat{\Pi}^{(n)}(\xi_n) = \hat{\Pi}_{\phi=1}^{(n)}(\xi_n)
$
for the normalized variable 
$
\xi_n = \delta u_n / \dbra (\delta u_n)^2 \dket^{1/2}
$,
whereas the PDF's $\hat{\Lambda}^{(n)}(\omega_n)$ 
of the pressure fluctuations and of the fluid particle accelerations are given by
the common formula
$
\hat{\Lambda}^{(n)}(\omega_n) = \hat{\Pi}_{\phi=2}^{(n)}(\omega_n)
$
for the normalized variable 
$
\omega_n = \delta p_n / \dbra (\delta p_n)^2 \dket^{1/2}
$.

\section{Log-normal model}

In the log-normal model \cite{Oboukhov62,K62,Yaglom}, one consider the ratio
$
\epsilon_n/\epsilon_{n-1}
$
$(n=1,2,\cdots)$
as independent stochastic variables, 
and apply for $n\gg 1$ the central limit theorem to the summation of their logarithms,
$
(n \sigma^2)^{-1/2} \sum_{j=1}^n \ln (\epsilon_j/\epsilon_{j-1}) 
= (n \sigma^2)^{-1/2} \ln (\epsilon_n/\epsilon)
= \sqrt{n} \sigma^{-1} (1-\alpha) \ln \delta
\label{def of variable}
$,
to have the Gaussian distribution function
$
P^{(n)}(\alpha) = (n/2\pi \sigma^2)^{1/2} 
\me^{- n(\alpha - \alpha_0 )^2/2\sigma^2}
\label{dist func for log-normal}
$
for the range $-\infty < \alpha < \infty$.
Here, we used the scaling relation between $\epsilon_n$ and $\alpha$.
Then, we have the multifractal spectrum and the mass exponent in the forms
$
f(\alpha) = 1 - (\alpha -\alpha_0)^2/2\sigma^2 \ln \delta
\label{f-alpha BG}
$
and 
$
\tau(\bar{q}) = 1-\alpha_0 \bq + \bq^2 \sigma^2 (\ln \delta)/2
$,
respectively.
We see that
$
a_{3\bq} = 1
$.

The dependence of the parameters $\alpha_0$ and $\sigma$ on 
the intermittency exponent $\mu$ is determined 
with the help of the two independent equations, i.e.,
the energy conservation:
$
\bra \epsilon_n/\epsilon \ket = 1
\label{cons of energy}
$ (equivalently, $\tau(1) = 0$), and
the definition of the intermittency exponent $\mu$:
$
\bra \epsilon_n^2/\epsilon^2 \ket 
= \delta_n^{-\mu}
\label{def of mu}
$ (equivalently, $\mu=1+\tau(2)$).
Here, $\epsilon$ is the energy input rate to the largest eddies.
The parameters are specified by means of $\mu$ as
$
\alpha_0 = 1 + \mu/2
$ and 
$
\sigma^2 = \mu/\ln \delta
$.
Then, we have 
$
f(\alpha) = 1 - (\alpha - \alpha_0)^2/2\mu
\label{f-alpha BG final}
$
and
$
\tau(\bq) = (1-\bq)D_{\bq}
$
with the generalized dimension 
$
D_{\bq}= 1-\mu \bq /2
$,
which are the same as derived in \cite{Meneveau87b}.
We know that 
$
\alpha_{\bq}=\alpha_0 - \mu \bq
$.

The explicit form of the PDF (\ref{PDF large phi}) 
for $\xi_n^* \leq \vert \xi_n \vert <\infty$ is found to be
\be
\hat{\Pi}_\phi^{(n)}(\xi_n)
= \bar{\Pi}_\phi^{(n)} \frac{\bar{\xi}_n}{\vert \xi_n \vert}
\exp \left[- \frac{\left(3 \ln \vert \xi_n / \xi_{n,0} \vert\right)^2}{
2 \phi^2 \mu \vert \ln \delta_n \vert} \right].
\label{PDF large log normal}
\ee
The connection point is given by 
$
\alpha^{*}= \alpha_0 -(\sqrt{3}-1)\phi \mu/3
$.

\section{$P$ model}

The distribution function $P^{(n)}(\alpha)$ for 
the $p$ model \cite{Meneveau87a,Meneveau87b} 
is specified based on the binomial multiplicative process 
in the form \cite{AA}
$
P^{(n)}(\alpha) = (Z_0^{(n)})^{-1}[2y^y (1-y)^{1-y} ]^{-n}
$
with
$
y = y(\alpha) = [\alpha + \log_2(1-p)]/\log_2[(1-p)/p]
$
and 
the partition function
$
Z_0^{(n)} = \sqrt{\pi/2n} \log_2[(1-p)/p]
$
for $n \gg 1$.
The multifractal spectrum is given by
$
f(\alpha) 
=-\{y(\alpha) \log_2 y(\alpha) +[1-y(\alpha)] \log_2 [1-y(\alpha)]\}
$,
which leads to the mass exponent 
$
\tau(\bq) = \log_2 [p^{\bq} + (1-p)^{\bq}]
$.
We have
$
f^{\prime \prime}(\alpha_{\bq}) = 
-\ln2 /\{\ln[(1-p)/p]\}^2 y_{\bq}(1-y_{\bq})
$
with 
$
y_{\bq}= y(\alpha_{\bq}) = p^{\bq}/[p^{\bq} + (1-p)^{\bq}]
$,
$
\alpha_{\bq} = - [p^{\bq} \log_2 p + (1-p)^{\bq}\log_2 (1-p)]/[p^{\bq}+(1-p)^{\bq}]
\label{alpha bq}
$,
and then we obtain
$
a_{3\bq} =2[p^{\bq} (1-p)^{\bq}]^{1/2}/[p^{\bq} +(1-p)^{\bq}]
$.
We see that, for $p > 1/2$,
$
\alpha_{\rm min} = - \log_2 p
$,
$
\alpha_{\rm max} = - \log_2 (1-p)
$.

The dependence of $p$ on the intermittency coefficient $\mu$ is derived 
through its definition $\mu = 1 + \tau(2)$ to give
$
p=(1+\sqrt{2^\mu-1})/2
$.

The explicit form of the PDF (\ref{PDF large phi})
for $\xi_n^* \leq \vert \xi_n \vert \leq \xi_n^{\rm max}$ reduces to 
\bea
\! \! \! \! \! \! \!
\hat{\Pi}_\phi^{(n)}(\xi_n)
= \bar{\Pi}_\phi^{(n)} \frac{\bar{\xi}_n}{\vert \xi_n \vert}
\left\{ \left[1 +2\left(y-y_0 \right)\right]^{1+2(y-y_0)}
\left[1 -2\left(y-y_0 \right)\right]^{1-2(y-y_0)} \right\}^{-n/2}
\eea
with 
$
y-y_0 = 3\ln (\vert \xi_n \vert/\vert \xi_{n,0} \vert)/
\phi \{\log_2 [(1-p)/p]\} \vert \ln \delta_n \vert
$.

\section{Harmonious representation}

In the harmonious representation of MFA, we adopt 
the Tsallis-type distribution function \cite{AA1,AA4}
$
P^{(n)}(\alpha) = ( Z_0^{(n)} )^{-1} \{ 1  - 
[ \left(\alpha - \alpha_0\right)/\itDelta \alpha ]^2 \}^{n/(1-q)}
\label{Tsallis prob density}
$
with 
$
(\itDelta \alpha)^2 = 2X/[(1-q) \ln 2]
$
and the appropriate partition function $Z_0^{(n)}$.
Here, $q$ is the entropy index introduced in the definitions of 
the R\'enyi \cite{Renyi} and the Tsallis entropies \cite{Tsallis88,Havrda-Charvat}.
This distribution function provides us with the multifractal spectrum
$
f(\alpha) = 1 + (1-q)^{-1} \log_2 [ 1 - (\alpha - \alpha_0 )^2/(\Delta \alpha )^2 ]
\label{Tsallis f-alpha}
$
which, then, produces the mass exponent
$
\tau(\bq) = 1-\alpha_0 \bq + 2X\bq^2 (1+ C_{\bq}^{1/2} )^{-1}
+ (1-q)^{-1} [1-\log_2 (1+ C_{\bq}^{1/2} ) ]
\label{tau}
$
with
$
{C}_{\bq}= 1 + 2 \bq^2 (1-q) X \ln 2
\label{cal D}
$.
We have
$
f^{\prime \prime}(\alpha_{\bq}) = - C_{\bq}^{1/2} (1+C_{\bq}^{1/2})/2X
$
with
$
\alpha_{\bq} = \alpha_0 - 2\bq X /(1+C_{\bq}^{1/2})
$,
hence
$
a_{3\bq} = \{ 2/[C_{\bq}^{1/2} ( 1+ C_{\bq}^{1/2})] \}^{1/2}
$.
We see that 
$
\alpha_{\rm min/max} = \alpha_0 \mp \Delta \alpha
$.

The dependence of the parameters $\alpha_0$, $X$ and $q$ on 
the intermittency exponent $\mu$ is determined, 
self-consistently, with the help of the three independent equations, i.e.,
the energy conservation: $\tau(1) = 0$, 
the definition of the intermittency exponent $\mu$: $\mu=1+\tau(2)$,
and the scaling relation:\cite{scaling relation}
$
1/(1-q) = 1/\alpha_- - 1/\alpha_+
\label{scaling relation}
$
with $\alpha_\pm$ satisfying $f(\alpha_\pm) =0$. 
Within the range $0.13 \leq \mu \leq 0.40$ where most of the experiments are covered, 
the self-consistent equations are solved, numerically, to give \cite{AA8}
$
\alpha_0 = 0.9989 + 0.5814 \mu
$,
$
X = 1.198 \mu
$
and
$
q = -1.507 + 20.58 \mu - 97.11 \mu^2 + 260.4 \mu^3 - 365.4 \mu^4 + 208.3 \mu^5
$.
It is interesting to compare the first two with the corresponding equations for 
the log-normal model ($q=1$), i.e., $\alpha_0 = 1 + 0.5 \mu$ and $X=\mu$.

The explicit form of the PDF (\ref{PDF large phi})
for $\xi_n^* \leq \vert \xi_n \vert \leq \xi_n^{\rm max}$ turns out to be
\be
\hat{\Pi}_\phi^{(n)}(\xi_n)
= \bar{\Pi}_\phi^{(n)} \frac{\bar{\xi}_n}{\vert \xi_n \vert}
\left[1 - \frac{1-q}{n}\ 
\frac{\left(3 \ln \vert \xi_n / \xi_{n,0} \vert\right)^2}{
2 \phi^2 X \vert \ln \delta_n \vert} \right]^{n/(1-q)}.
\label{PDF large}
\ee

\section{Analysis of experiments}

The scaling exponents  $\zeta_m$ of velocity structure function 
reported by Gotoh et al.\ \cite{Gotoh02}
are analyzed by the method of the least squares (MLS) with the theoretical formulae 
of the harmonious representation, of the log-normal model and of the $p$ model,
giving, respectively, the values of the intermittency exponent $\mu =$ 
0.240, 0.217 and 0.249.

\begin{figure}
\begin{center}
\includegraphics[width=13cm]{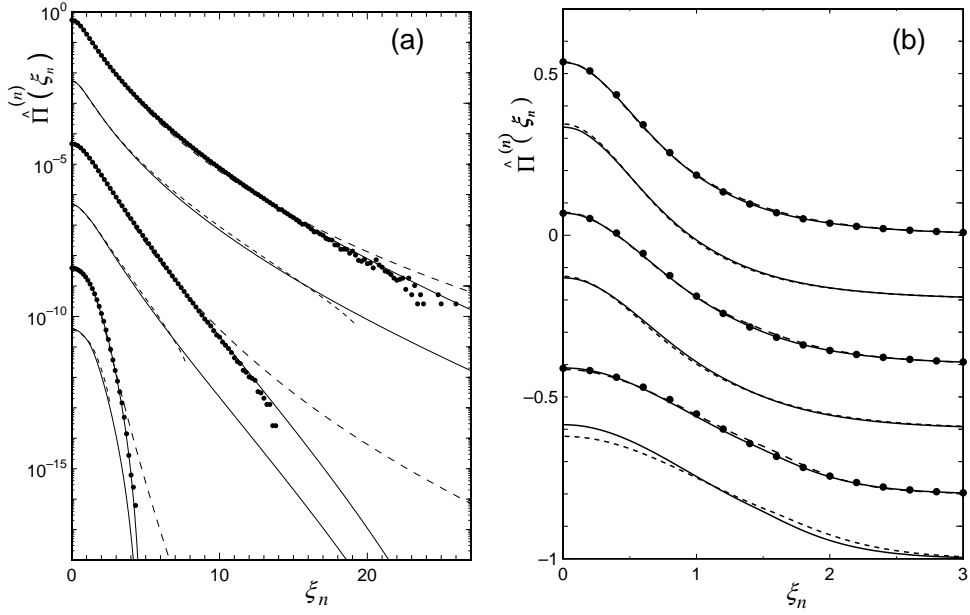}
\end{center}
\caption{Analyses of the PDF's of the velocity fluctuations (closed circles)
for three different measuring distances, 
observed by Gotoh et al.\ at $R_\lambda = 380$, with the help of
the PDF's $\hat{\Pi}^{(n)}(\xi_n)$ by the harmonious representation (solid line)
and by the log-normal model (dashed line)
are plotted on (a) log and (b) linear scales.
The PDF's by the $p$ model (dotted line) are compared with the PDF's by
the harmonious representation (solid line).
Comparisons are displayed in pairs. The solid lines in each set of pairs are the same.
For better visibility, each PDF is shifted by $-2$ unit in (a) and 
by $-0.2$ in (b) along the vertical axis.
Parameters are given in the text.
\label{Gotoh fluc 380 log-normal}}
\end{figure}
In Fig.~\ref{Gotoh fluc 380 log-normal}, 
the PDF's of the velocity fluctuations (closed circles) 
measured by Gotoh et al.\ in their DNS at $R_\lambda = 380$ \cite{Gotoh02} 
for three different measuring distances,
$r/\eta = \ell_n/\eta =$ 2.38, 19.0, 1220 from the top set of pairs to the bottom set,
are analyzed with the help of
the PDF's $\hat{\Pi}^{(n)}(\xi_n)$ of the harmonious representation (solid line)
and of the log-normal model (dashed line).
Here, $\eta = (\nu^3/\epsilon )^{1/4}$ represents the Kolmogorov scale.
The DNS data points are symmetrized by taking averages of 
the left and the right hand sides data.
In each set, a comparison of the PDF of the harmonious representation (solid line) 
with the PDF of the $p$ model (dotted line) is given. 
For the harmonious PDF's (solid line), 
$q=0.391$ ($\mu = 0.240$), and, from the top set to the bottom set, $(n,\ q') =$
(20.7,\ 1.60), (13.6,\ 1.50), (6.10,\ 1.20),
$
\xi_n^* = 1.10, 1.23, 1.43
$
($\alpha^* = 1.07$)
and
$
\xi_n^{\rm max} = 204, 38.2, 6.63
$.
For the PDF by the log-normal model (dashed line), 
from the top set to the bottom set, $(n,\ q') =$ 
(21.5,\ 1.70), (13.0,\ 1.63), (5.00,\ 1.24) and
$
\xi_n^* = 1.19, 1.34, 1.51
$
($\alpha^* = 1.06$).
For the PDF by the $p$ model (dashed line), 
from the top set to the bottom set, $(n,\ q') =$
(21.0,\ 1.60), (13.0,\ 1.62), (5.50,\ 1.20), 
$
\xi_n^* = 1.06, 1.07, 1.55
$
($\alpha^* = 1.08$)
and
$
\xi_n^{\rm max} = 19.3, 7.70, 3.31
$.
The tail of PDF of the $p$ model stops at $\xi_n^{\rm max}$ which is smaller than 
the maximum value of measured data point for each measuring distance.
Note that every PDF's are plotted on (a) log and (b) linear scales.

\begin{figure}
\begin{center}
\includegraphics[width=13cm]{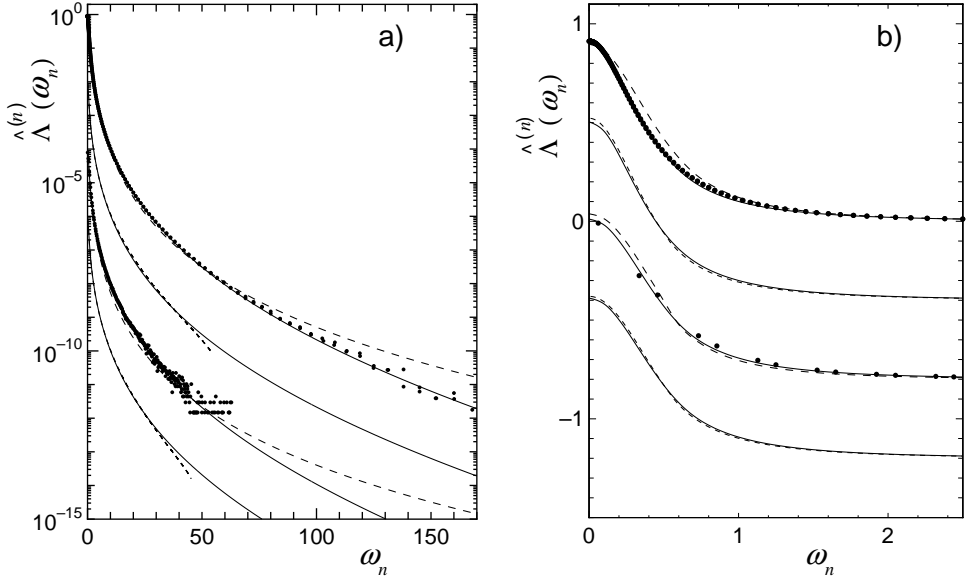}
\end{center}
\caption{Analyses of the PDF's of {\it fluid particle accelerations},
measured by Gotoh et al.\ at $R_\lambda = 380$ (circles in the top set) 
and by Bodenschatz et al.\ at $R_\lambda = 690$ (circled in the bottom set),
by means of the PDF's $\hat{\itLambda}^{(n)}(\omega_n)$ 
by the harmonious representation (solid line) 
and by the log-normal model (dashed line)
are plotted on (a) log and (b) linear scales.
The PDF's by the $p$ model (dotted line) are compared with the PDF's by
the harmonious representation (solid line).
Results are displayed in pairs. The solid lines in each set of pairs are the same.
For better visibility, each PDF is shifted by $-2$ unit in (a) and 
by $-0.4$ in (b) along the vertical axis.
Parameters are given in the text.
\label{Gotoh Boden accel log-normal}}
\end{figure}
In Fig.~\ref{Gotoh Boden accel log-normal}, the PDF's of 
the {\it fluid particle accelerations} (closed circles) reported 
by Gotoh et al.\ at $R_\lambda = 380$ (top set) \cite{Gotoh pressure}
and by Bodenschatz et al.\ at $R_\lambda = 690$ (bottom set) \cite{EB01a,EB01b,EB02comment}
are analyzed with the PDF's $\hat{\itLambda}^{(n)}(\omega_n)$
of the harmonious representation (solid line)
and of the log-normal model (dashed line).
The measured data points both on the left and right hand sides 
of the PDF's are shown altogether on one side by closed circles in the figure.
In each set, a comparison of the PDF of the harmonious representation (solid line)
with the PDF of the $p$ model (dotted line) is given.
For the harmonious PDF's (solid line), from the top set of pairs to the bottom set, 
$q=$0.391, 0.391 ($\mu =$ 0.240, 0.240), 
$(n,\ q')=$(17.5, 1.70), (17.1, 1.45)
$\omega_n^* =$ 0.622, 0.605
($\alpha^* =$ 1.01, 1.01), and
$\omega_n^{\rm max} =$ 2530, 2040.
For the PDF by the log-normal model (dashed line), from the top set to the bottom set, 
$(n,\ q')=$ (17.0, 1.50), (18.5, 1.04),
$\omega_n^* =$ 0.644, 0.558
($\alpha^* =$ 1.00, 1.00), and
$\omega_n^{\rm max} =$ 49.6, 76.6.
For the PDF by the $p$ model (dotted line), from the top set to the bottom set, 
$(n,\ q')=$ (19.0, 1.50), (18.5, 1.20)
$\omega_n^* =$ 0.539, 0.547
($\alpha^* =$ 1.01, 1.01), and
$\omega_n^{\rm max} =$ 54.1, 45.2.
Note that every PDF's are plotted on (a) log and (b) linear scales.

The values of $n$ for all the PDF's are determined by MLS by adjusting 
the integrand $\xi_n^{4} \Pi_\phi^{(n)} (\xi_n)$ of the fourth moment 
both of data and of the theories.
The values of $q^\prime$ for all the PDF's are obtained by MLS by fitting
the theoretical PDF's at the center part, $\vert \xi_n \vert \leq \xi_n^*$, 
with the observed PDF's.

\section{Discussions}

As was shown in this paper, the harmonious representation within MFA
provides the highest accuracy in analyzing observed PDF's. 
This high accuracy allows us to extract some information about the dynamics underlying
turbulence.
For example, through the analysis of the velocity fluctuations
in Fig.~\ref{Gotoh fluc 380 log-normal}, we extracted 
the dependence of $q^\prime$ on $r/\eta$ as \cite{AA12}
$
q^\prime = 1.71 - 0.05 \log_2 (r/\eta)
$.
We then obtain the diffusion coefficient for the fluctuations of 
magnitude $\vert \xi_n \vert$ ($\leq \xi_n^*$) in the form
$
D_{\phi}(\xi_n)/D_{\phi}(0)= 1 
+ [0.71 - 0.05 \log_2 (r/\eta) ] 
[(\phi + 3 f^\prime(\alpha^*))/2\phi] (\xi_n/\xi_n^*)^2
$.
Note that this is dependent on the distance $r$ of two measuring points. 
In deriving this, we assumed that the drift term is linear since the magnitude of
the fluctuations in the center part of the PDF is smaller 
than the order of their standard deviations ($\vert \xi_n \vert \leq 1$), 
and, therefore, used the formula for anomalous diffusion given in \cite{Kaniadakis00}.
The investigation to this direction is in progress, and will be reported elsewhere.

\section*{Acknowledgements}

The authors would like to thank Prof.~C.~Tsallis for his continuous encouragement,
and Dr.~A.M.~Reynolds for the fruitful discussions with him.
The authors are grateful to Prof.~E.~Bodenschatz and Prof.~T.~Gotoh
for the kindness to show their data prior to publication.


\begin{thebibliography}{00}
\bibitem{scaling relation} The scaling relation is a generalization of the one derived first in
		\cite{Costa,Lyra98} to the case where the multifractal spectrum
		has negative values \cite{AA1}.
\end{thebibliography}
\end{document}